## Median–Radial Function: A Robust, Covariance-Free Framework and Applications

**Elsayed A. H. Elamir**
Department of Management and Marketing, College of Business Administration, Kingdom of Bahrain
Email: shabib@uob.edu.bh

**Abstract**
A median-radius framework for assessing centrality in multivariate data using median distances is proposed. Based on the proposed framework, scale-invariant measure of radial concentration is defined with particular attention given to multimodal settings where nonconvexity of the median radius field may generate multiple concentration regions. This motivates interpreting the functional as a robust concentration landscape. Fundamental properties of the proposed functional including continuity, Lipschitz and existence of minimizers are established. The derivative and generalized gradient structure are obtained. Monte Carlo simulation and applications demonstrate the ability of the framework to describe asymmetric and multimodal structures without covariance estimation or inversion and provides a complementary covariance-free framework for investigating multivariate centrality, concentration and local directional structure.

ORCID: 0000-0002-9430-072X

**Conflict of interest:** The author does not have any conflict of interest.
**Financial support**: This research did not receive any specific grant from funding agencies in the public, commercial, or not-for-profit sectors.
**AI Tool**: ChatGPT and copilot are used for improving language, readability and presentation

# 1 Introduction

Measures of multivariate centrality are traditionally based on either distance or depth. Among distance-based approaches, the Mahalanobis distance ("Reprint Of," 2018) which incorporates covariance structure and yields affine-invariant elliptical geometry. While optimal under Gaussian assumptions, it is highly sensitive to outliers and becomes unreliable in high-dimensional settings due to instability of covariance estimation (Cheng et al., 2024; Huber & Ronchetti, 2009; Park et al., 2024). While these methods are optimal under Gaussian assumptions, they are highly sensitive to outliers, heavy-tailed or skewed distributions and become unstable in high-dimensional settings (Rousseeuw & Croux, 1993; Schwank & Drton, 2026). To improve robustness, a variety of estimators of scatter have been developed, including the minimum covariance determinant (MCD) and S-estimators (Rousseeuw & Croux, 1993). These lead to robust Mahalanobis distances with high breakdown points. These approaches can substantially improve resistance to contamination and provide robust alternatives to classical Mahalanobis distance. Nevertheless, they dependent on scatter estimation and matrix inversion which can present computational and statistical difficulties when the dimension is large relative to the sample size (García-Pérez, 2023; Hunter, 2026; Ley & Verdebout, 2017). Furthermore, their associated distance contours retain an elliptical structure that may not adequately represent asymmetric or multimodal distributions.

Median-based statistics offer another way to robust measurement of location and dispersion. The median absolute deviation (MAD) is a well-established robust measure of univariate scale (Ku & Han, 2008; Rousseeuw & Leroy, 1987). In multivariate analysis, the geometric median extends the notion of central location by minimizing expected Euclidean distance. However, this construction differs from minimizing the median of Euclidean distances. The latter criterion emphasizes the radius required to capture at least half of the probability mass and therefore provides a distinct characterization of multivariate concentration (Johnson & Wichern, 2007; Maronna et al., 2018).

Motivated by this distinction, we consider the median-radius functional (MRF), also referred to as the moving median absolute deviation (MMAD), defined for a random vector $X \in \mathbb{R}^p$ and a reference point $v \in \mathbb{R}^p$ by

$$G_F(v) = \mathrm{Med}\|X - v\|,$$

Where the median is interpreted as the lower half quantile of the distance distribution.

The MRF provides a geometric description of the directional organization of data near the median radius boundary. By examining the unit directions associated with observation in an active shell, a shell-direction convex hull and corresponding directional imbalance measure are constructed. These quantities summarize local geometric structure without requiring third-order moments or covariance inversion. In contrast to classical multivariate skewness measures, which describe global distributional asymmetry, shell imbalance characterizes the directional organization of observations around a specified reference point. It is therefore intended as a complementary geometric diagnostic rather than a direct replacement for conventional skewness measures (Deb & Sen, 2023; Ley & Verdebout, 2017).

The main contribution of this study is the development and investigation of a covariance-free median-radius framework for multivariate centrality, concentration and directional structure. We establish fundamental properties of the median-radius functional including continuity, Lipschitz behaviour and existence of minimizers and transformation properties. We further investigate its normalized form and its geometric interpretation under nonconvex and multimodal settings. The proposed framework is evaluated through comparisons with classical and robust Mahalanobis distances, geometric analyses of shell directions and Monte Carlo experiments examining the sampling stability of shell imbalance relative to established multivariate skewness measures under heavy-tailed distributions. A real-data illustration is also used to demonstrate the interpretation of active-shell geometry and directional convex hulls. The remainder of the paper is organized as follows. Section 2 introduces the median-radius functional and investigates its principal theoretical properties. Section 3 extends the framework to multivariate distributions and establishes its transformation properties. Section 4 presents numerical and empirical applications that includes comparisons with classical and robust Mahalanobis distances, the geometric behaviour of shell directions and the sampling stability of shell skewness relative to classical measures. Section 5 is devoted to conclusion.

## 2 Median-radius function

The median-radius function (MRF) is defined for any $v \in \mathbb{R}$ by

$$G_X(v) = \text{Med}|X - v|,$$

Where Med denotes the lower median quantile. It provides a robust, median-based measure of the dispersion of $X$ around the reference point $v$ rather than around a fixed location (Elamir, 2025). It generalizes the classical median absolute deviation (MAD) by allowing the centre to vary across the real line.

**Definition 1.** (Quantile representation of $G(v)$)
Let $X$ be a real-valued random variable with cumulative distribution function $F$. For any $v \in \mathbb{R}$ and $r \geq 0$, define the centred probability mass function

$$M_X(v, r) := \mathbb{P}(|X - v| \leq r) = F(v + r) - F(v - r).$$

The median–distance functional

$$G_F(v) = \text{Med}(|X - v|),$$

admits the quantile representation

$$G_F(v) = Q_{0.5}(|X - v|) = \inf\{r \geq 0 : M_X(v, r) \geq 0.5\}.$$

Thus, $G(v)$ is the smallest radius centred at $v$ that captures at least half of the probability mass. In particular, the differentiability and smoothness properties of $G(v)$ depend on how the distribution allocates probability mass near the boundary points $v \pm G(v)$.

**Theorem 1** (Lipschitz continuity)
Let $X$ be a real-valued random variable and $G_F(v) = \text{Med}(|X - v|)$. For any $v, u \in \mathbb{R}$,

$$|G_F(v) - G_F(u)| \leq |v - u|.$$

Hence $G_F$ is $1-$Lipschitz and therefore continuous.
*Proof.* See Appendix.

**Theorem 2** (Coercivity and existence)
For any probability distribution $F$ on $\mathbb{R}$,

$$G_F(v) \longrightarrow \infty, \quad as\ |v| \to \infty.$$

Consequently, since $G_F$ is continuous, the minimizer set

$$C_F = \arg\min_{\text{v}} G_F(v),$$

is nonempty and compact.
*Proof.* see Appendix

## 2.1 Median-radius center

**Definition 2** (Median-radius centre)
Let

$$C_F = \arg\min_{v\in\mathbb{R}} G_F(v),$$

and

$$g_F = \min_{\mathrm{v}} G_F(v),$$

The set $C_F$ is called the median-radius centre set. When the minimizer is unique, its element is denoted by $C_F$. In some regular symmetric or unimodal settings, the median-radius centre may coincide with the ordinary median. However, this identity is not universal. The median-radius centre is defined by minimizing the median distance while the ordinary median is defined through probability balance.

**Proposition** (Nonconvexity in general)
The median-radius functional

$$G_F(v) = \mathrm{Med}(|X - v|),$$

need not be convex even for a continuous distribution $F$. Multimodal distributions may produce multiple regions of relatively small median radius by regions with larger values.
*Proof.* See Appendix.

**Definition 3** (Standardized median-radius functional).
Assuming $g_F > 0$, let

$$H_F(v) = \frac{G_F(v)}{g_F}.$$

Thus, $H(v)$ measures the median absolute deviation from $v$, normalized by its minimum median radius $g_F$. Therefore, $H_F(v) \geq 1$ and $H_F(v) = 1 \Leftrightarrow v \in C_F$. Furthermore, from Lipschitz continuity of $G(v)$,

$$|H_F(v) - H_F(w)| \leq \frac{|v - u|}{g_F}.$$

The function $H(v)$ is a standardized version of the median–radius functional that preserving its geometric and minimizers while providing a scale-free measure of relative radial concentration.

**Theorem 3** (derivative of $G(v)$)
Assume $X$ is one-dimensional with a continuously differentiable cumulative distribution function $F$ and density $f$. Suppose that $G(v)$ is locally uniquely determined by $F\big(v + G(v)\big) - F\big(v - G(v)\big) = 0.5$ and that $f\big(v - G(v)\big) + f\big(v + G(v)\big) > 0$. Then $G(v)$ is differentiable at $v$, with

$$G'(v) = \frac{f\big(v - G(v)\big) - f\big(v + G(v)\big)}{f\big(v - G(v)\big) + f\big(v + G(v)\big)}.$$

*Proof.* see Appendix.

**Corollary** (Bounded derivative)
Under the assumptions of Theorem 3. The range of $G'(v)$ is

$$|G'(v)| \leq 1.$$

This is consistent with the global 1-Lipschitz property established in Theorem 1.

*Proof.* See Appendix.

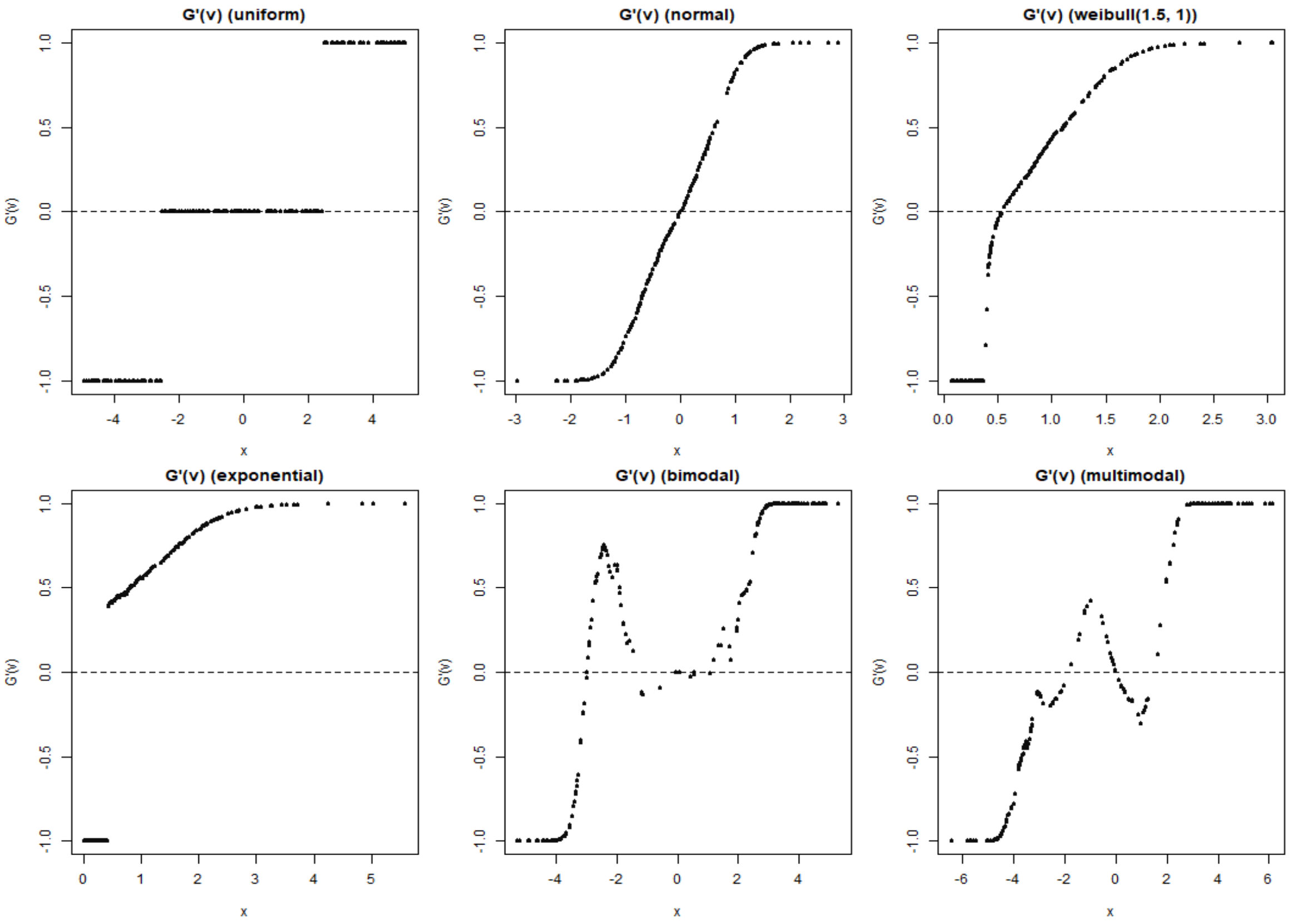

Fig 1. Derivative profiles of MRF $G'(v)$ for different distributions using $n = 200$.

Figure 1 illustrates the behaviour of the derivative $G'(v)$ for six distribution (uniform (-5,5), normal (0,1), Weibull (1.5,1), exponential (1), bimodal normal ((-3,3), (1,1)) and multimodal normal ((-4,0,4), (1,1,1)). For each distribution, $n = 200$ detrmisistic observations were generated using $x_i = Q((i - 0.5)/n)$. The results reveal distinct patterns associated with distributional shape. For the uniform distribution, the derivative exhibits piecewise-constant behaviour. The normal distribution produces a smooth, approximately symmetric transition from $-1$ to 1 with a zero crossing near the distribution centre. The Weibull and exponential distributions display asymmetric derivative profiles, consistent with their right-skewed structures. In contrast, the bimodal and multimodal distributions reflecting the presence of separated concentration regions where they exhibit more complex patterns, including changes in sign and local oscillations.

## 2.2 Clarke generalized gradients.

Since the median-radius function $G_F(v)$ is globally Lipschitz, it is differentiable almost everywhere. At points where ordinary differentiability fails; its local behaviour can be characterized using the Clarke generalized gradient. For a locally Lipschitz function $G: \mathbb{R} \to \mathbb{R}$, the Clarke generalized gradient at $v$ is defined as

$$\partial_C G_F(v) = \text{conv}\left\{\lim_{k\to\infty} G'(v_k): v_k \to v, G_F \text{ differentiable at } v_k\right\}.$$

Where the convex hull is taken over all existing limits of derivatives along sequences converging to $v$ (Clarke, 1990; Clason & Valkonen, 2026).
In one dimension, $\partial_C G_F(v)$ is a nonempty compact interval. If $G$ is piecewise continuously differentiable in a neighbourhood of $v$, with finite one-sided derivative $G'_-(v)$ and $G'_+(v)$ then

$$\partial_C G_F(v) = [\min\{G'_-(v), G'_+(v)\}, \max\{G'_-(v), G'_+(v)\}].$$

At points where $G$ is continuously differentiable in a neighbourhood of $v$,

$$\partial_C G_F(v) = \{G'(v)\}.$$

Since $G(v)$ is globally 1-Lipschitz, every generalized gradient satisfies $\partial_C G_F(v) \subseteq [-1,1]$.
Furthermore, a necessary condition for $v^*$ to be local minimum is $0 \in \partial_C G_F(v^*)$.
However, $G(v)$ need not be convex, this condition is not sufficient to establish a local or global minimum. The Clarke generalized gradient is therefore appropriate for investigating the local geometry of the median-radius function that includes its stationary points and nondifferentiable regions. Unlike the convex subdifferential, it remains applicable when the median-radius function is nonconvex (Boyd & Vandenberghe, 2004).

# 3 Multivariate Median-radius function

Let $X$ be a random vector in $\mathbb{R}^d$with distribution $F$. The univariate function

$$G(v) = \text{Med}(|X - v|)$$

admits a natural extension to the multivariate setting by replacing the absolute value with a norm. Throughout, $\|\cdot\|$ denotes the Euclidean norm unless otherwise specified.

**Definition 4.** (Multivariate median–distance functional).
For any $v \in \mathbb{R}^d$, define

$$G_F(v) = \text{Med}\|X - v\| = \inf\{r \geq 0: P(\|X - v\| \leq r) \geq 0.5\},$$

Where Med denotes the lower median quantile of the median of the distribution of the random variable $\|X - v\|$. Equivalently, $G_F(v)$ is the smallest radius such that the closed Euclidean ball

$$B(v, G(v)) = \{x \in R^p: \|x - v\| \leq G_F(v)\},$$

contains at least one half of the probability mass of $F$.
Thus $G(v)$ represents the median radius of the data distribution when viewed from the reference point $v$. The median-radius centre set is defined as

$$C_F = \arg\min_{v \in \mathbb{R}^p} G_F(v),$$

The multivariate median-radius functional is continuous and coercive as established by Theorem 1 and 2 whose arguments extend directly to Euclidean space. However, it need not be convex or quasiconvex. This flexibility allows the median-radius landscape to represent multiple regions of relatively high concentration, potentially separated by regions requiring larger radii to contain half of probability mass.

**Theorem 4** (similarity equivariance)
Let $X$ be a random vector in $\mathbb{R}^p$ with distribution $F$ and define

$$G_X(v) = \text{Med}\|X - v\|,$$

Consider the transformed random vector $Y = a + bQX,$ where $a \in \mathbb{R}^{\mathrm{p}},\ b > 0,\ Q^T Q = I$ (orthogonal matrix).
Then, for any $v \in \mathbb{R}^p,$

$$G_Y(a + bQv) = bG_X(v).$$

Consequently, if $C_X = arg \min_{v \in \mathbb{R}^p} G_X(v), g_X = \min_{w \in \mathbb{R}^p} G_X(v).$ Then $C_Y = a + bQC_X,$ and $g_Y = bg_X,$
Furthermore, provided $g_X > 0$ the standardized median-radius functional

$$H_X(v) = \frac{G_X(v)}{g_X},$$

satisfies

$$H_Y(a + bQv) = H_X(v).$$

Thus, $G$ is equivariant under translations, rotations, reflections, and common positive scaling, while $H_F$ is invariant under these similarity transformations.
Note that the median-radius functional is not generally affine equivariant where theorem 4 establishes similarity equivariance.
In this article, covariance-free means that neither the definition nor the computation of $G(v)$ and $H(v)$ or the corresponding median-radius centrality requires the estimation or inversion of a covariance or scatter matrix. The proposed framework nevertheless relies on Euclidean geometry and is therefore invariant under similarity transformations rather than arbitrary non-singular affine transformations.

# 4 Applications

We investigate two applications of the proposed median–radius framework. The first compares the geometric behaviour of the median-radius functional with classical and robust Mahalanobis distances under different distribution settings. The second examines shell directional asymmetry through the directional organization of observations near the median-radius boundary. These applications illustrate how the proposed framework provides complementary information about multivariate concentration, asymmetry and distributional structure.

## 4.1 Comparison with Mahalanobis distance

We compare the proposed median-radius functional with classical and robust Mahalanobis distances, focusing on their underlying assumptions, geometric behaviour, sensitivity to distributional structure and applicability in high-dimensional settings.
The classical Mahalanobis distance is defined as

$$D_M(x) = \sqrt{(x - \mu)^{\top} \Sigma^{-1} (x - \mu)},$$

where $\mu$ and $\Sigma$ denote the mean vector and positive definite covariance matrix, respectively.
The Mahalanobis distance accounts for the covariance structure of the data and produces elliptical contours. However, its classical sample implementation is sensitive to outliers and may become unreliable when covariance estimation is unstable or when the sample covariance matrix is singular. To address the limitations, robust estimators of location and scatter have been developed, including the minimum covariance determinant (MCD), S-estimators, and the orthogonalized Gnanadesikan–Kettenring (OGK) estimator (Hubert et al., 2005; Maturo et al., 2025). The corresponding robust Mahalanobis distance is

$$D_R(x) = \sqrt{(x - \hat{\mu}_{\mathrm{R}})\, \hat{\Sigma}^{-1}(x - \hat{\mu}_{\mathrm{R}})}.$$

where $\hat{\mu}_R$ and $\hat{\Sigma}_R$ are robust estimates of location and scatter, respectively.
These methods can substantially reduce to contamination depending on the estimators and its tuning parameters. Despite these improvements, they continue to rely on scatter estimation and retain elliptical distance contours. In high-dimensional settings, where the number of variables is comparable to or exceeds the sample size, the estimated scatter matrix may be singular or difficult to estimate reliably.
On the other hand, the proposed median-radius functional

$$G(v) = Med\|X - v\|,$$

and its standardized version

$$H(v) = \frac{G(v)}{g_F},$$

depend directly on Euclidean distances and do not require covariance estimation or matrix inversion. Their contours are not restricted to elliptical shapes and may reflect more complex concentration patterns.

### 4.1.1 Bimodal distribution

Figure 2 compares the median-radius contours with classical and robust Mahalanobis-distance contours for a two-component Gaussian mixture. The first component is generated from a bivariate normal distribution with an identity covariance matrix, while the second component has a similar covariance structure and is centred at $(2,0)$.
The resulting distribution exhibits two horizontally separated concentration regions. The scatter plot illustrates the mixture structure while the marginal boxplots summarize the variation in each coordinate. The contours of $G(v)$ and $H(v)$ display non-elliptical shapes influenced by the spatial arrangement of the two components. In contrast, both classical and robust Mahalanobis distances produce elliptical contours determined by their respective estimates of location and scatter.
This example illustrates the ability of the median-radius functional to describe half-mass concentration through a geometry that is not restricted to covariance-based elliptical contours.

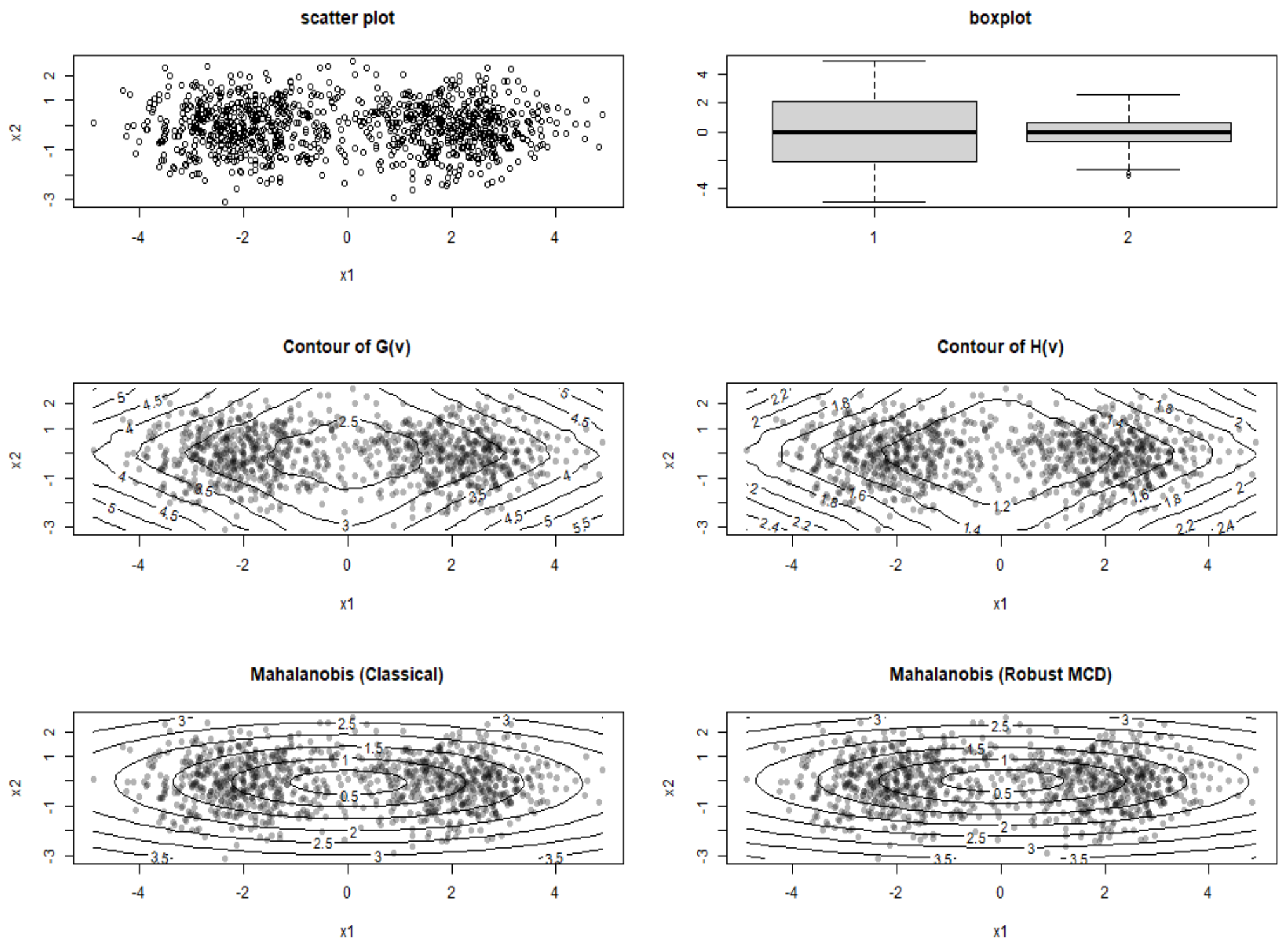

Fig 2. Comparison of median-radius contours and Mahalanobis distances under bimodal data.

### 4.1.2 Asymmetric distribution

Figure 3 presents a comparison under an asymmetric distribution generated by combining a normally distributed variable with an exponentially distributed variable. This construction produces a bivariate distribution that is approximately symmetric in one coordinate and positively skewed in the other. The contours of $G(v)$ and $H(v)$ reflect the asymmetric spatial distribution of observations especially the extension of the data cloud along the skewed direction.

In comparison, classical and robust Mahalanobis distances retain elliptical contours, even when robust estimates of location and scatter are employed. Although these distances account for dispersion and correlation, their elliptical geometry does not directly represent the asymmetric concentration pattern observed in this example. The results illustrate the flexibility of the median-radius functional in describing concentration patterns under skewed distributions.

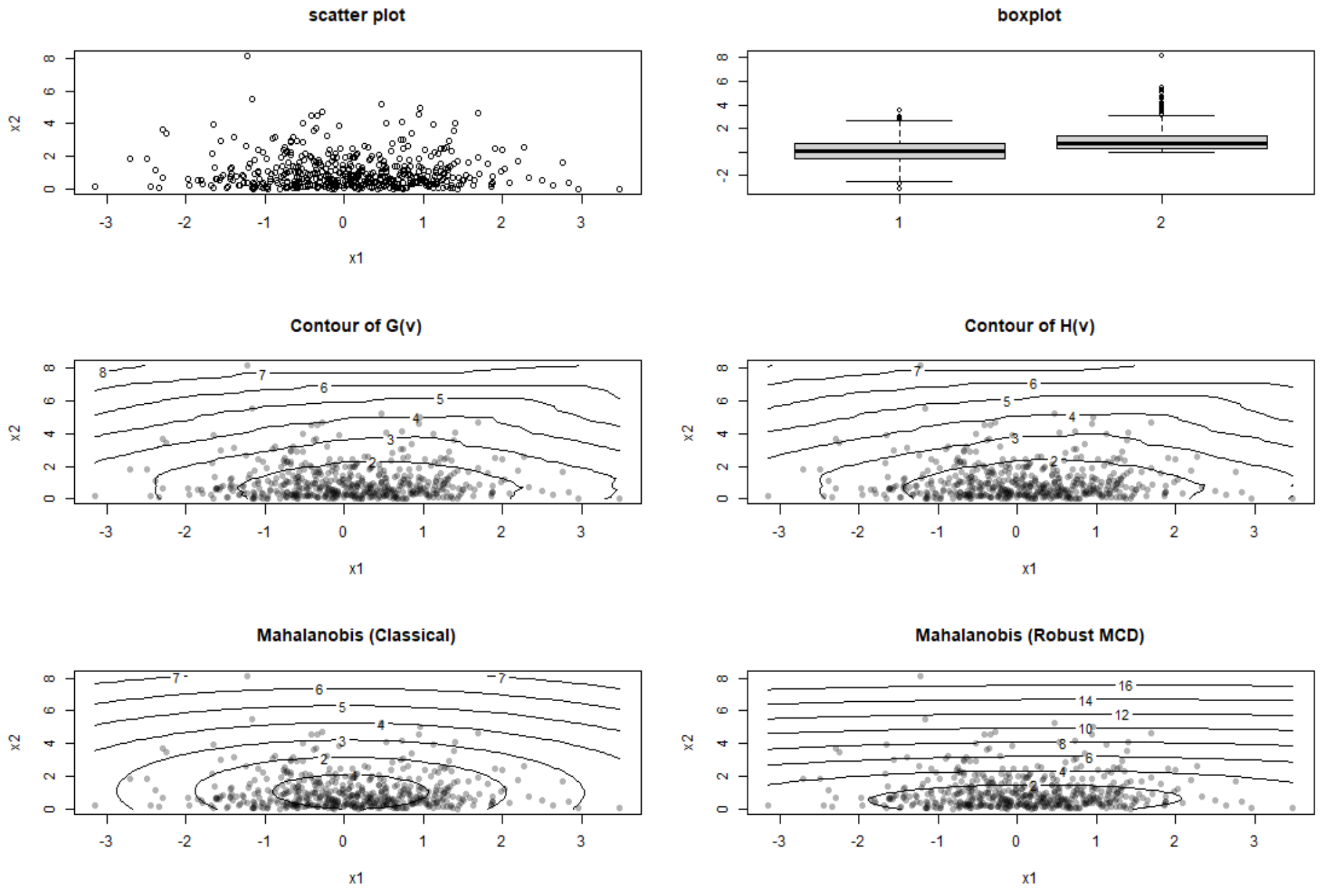

Fig 3. Comparison of median-radius contours and Mahalanobis distances under asymmetric data.

### 4.1.3 High-dimensional setting

Figure 4 illustrates a high-dimensional setting in which the number of variables exceeds the number of observations $p > n$. In this situation, the sample covariance matrix is singular and the classical Mahalanobis distance cannot be computed using the ordinary matrix inverse. Robust Mahalanobis procedures that also require inversion of a non-singular scatter matrix may encounter similar difficulties.

Figure 4 demonstrates this computational advantage by displaying the median-radius contours in a setting where the implemented Mahalanobis procedures encounter singularity problems. Although the median-radius functional remains mathematically well-defined in high dimensions, this does not automatically guarantee statistical robustness or improved discrimination. Its performance may still be influenced by distance concentration and other high-dimensional effects.

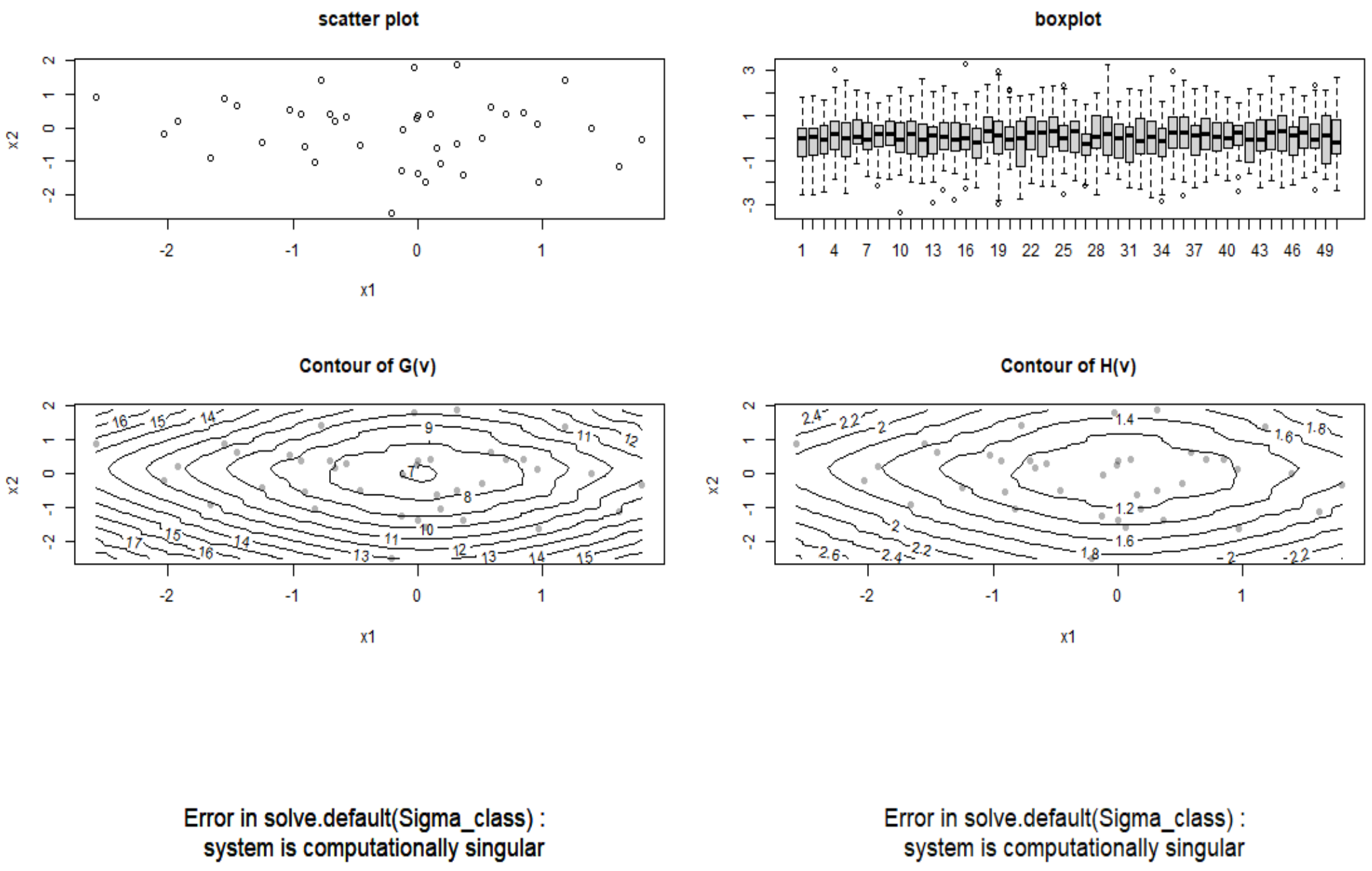

Fig 4. Median-radius in a high-dimensional setting ($p > n$).

Therefore, the comparisons demonstrate that the median-radius framework provides a covariance-free alternative for describing multivariate concentration. Its principal advantage is the ability to represent non-elliptical concentration landscapes without relying on covariance estimation. These properties complement rather than universally replace the classical and robust Mahalanobis distances.

### 4.1.4 Computational complexity

The proposed median-radius framework is computationally straightforward because it relies on Euclidean distances and median calculations rather than covariance estimation or matrix inversion. For a sample of $n$ observations in $p$ dimensions, evaluating the empirical median-radius functional

$$\hat{G}(v) = Med_{1\leq i\leq n}\|X_i - v\|,$$

at a fixed reference point $v$ requires $O(np)$ operations to compute the Euclidean distances. The sample median can be obtained in $O(n)$ expected time using a selection algorithm, giving an overall expected complexity of $O(np)$ per evaluation. Consequently, evaluating $\hat{G}$ at $m$ reference points require $O(mnp)$ operations. Identifying the active-shell observations and calculating the shell imbalance vector also require $O(np)$ operations for a fixed reference point. The overall cost of estimating the median-radius centre depends on the optimization algorithm and the number of function evaluations.

## 4.2 Shell-based directional measure

While the directions of observations near the median-radius boundary offer additional insight into the angular organization of the distribution, the median-radius functional provides information about the radial concentration of multivariate data, Motivated by the generalized gradient framework discussed in Section 2.2, we introduce a shell-based directional measure to examine local asymmetry around a reference point.
For a sample $X_1, \dots, X_n \in \mathbb{R}^p$, let

$$\hat{G}(v) = Med_{1\le i\le n} \|x_i - v\|,$$

Denote the empirical median-radius function. For any reference point $v \in \mathbb{R}^p$, the active shell may be defined as

$$A_\varepsilon(v) = \{i: \big| \|X_i - v\| - \hat{G}(v) \big| \le \varepsilon\},$$

Where $\varepsilon > 0$ is a tolerance parameter controlling the thickness of the shell. In practice, we use $\varepsilon = 0.10\ \hat{G}(v)$ corresponding to a tolerance of approximately 10% of the median radius.
For each $i \in A_\varepsilon(v)$, the unit directional vector can be defined as

$$s_i(v) = \frac{v - X_i}{\|v - X_i\|}.$$

Each vector represents the direction from a shell observation toward the reference point. These vectors are also the gradients, with respect to $v$, of the individual distance functions $\|X_i - v\|$ at points where $X_i \ne v$.
A natural shell skewness vector is defined as

$$\hat{\Gamma}_\in(v) = \frac{1}{|A(v)|} \sum_{i \in A(v)} s_i(v),$$

provided the active shell is nonempty and all included directions are defined.
Since each directional vector has unit norm $\|\hat{\Gamma}_\in(v)\| \le 1$, the scalar measure $S_\varepsilon(v) = \|\hat{\Gamma}(v)\|$ satisfies $0 \le S_\varepsilon(v) \le 1$.
The vector $\hat{\Gamma}_\in(v)$ and its magnitude $S_\in(v)$ provide complementary information about the directional organization of observations near the median-radius shell. In particular:

- $S_\varepsilon(v) \approx 0$ indicates substantial cancellation among shell directions and approximate directional balance.
- Larger values of $S_\varepsilon(v)$ indicate stronger directional imbalance among the selected shell observations.
- When $\hat{\Gamma}_\in(v) \ne 0$, its direction identifies the resultant orientation of the shell vectors.

Thus, the proposed measure characterizes shell directional asymmetry relative to the selected reference point. Although this information may be associated with distributional skewness, it may also reflect multimodality, anisotropy or the location of the reference point.

### 4.2.1 Geometric interpretation using the convex hull

The directional organization of the active shell can also be represented through the convex hull of its unit vectors. For a set of vectors $S = \{s_1, \dots, s_k\}$, the convex hull can be defined as

$$\text{conv}(S) = \left\{ \sum_{i=1}^{k} \lambda_i s_i : \lambda_i \ge 0, \sum_{i=1}^{k} \lambda_i = 1 \right\}.$$

Accordingly, define the empirical shell-direction hull as

$$S_\in(v) = \text{conv}\{s_i(v) : i \in A_\varepsilon(v)\}.$$

Since the vectors $s_i(v)$ lie on the unit sphere, their convex hull is a compact convex set contained in the closed unit ball, $S(v) \subseteq \{z \in R^p : \|z\| \leq 1\}$. Moreover, the shell skewness vector is an equal-weight convex combination of the shell directions, and therefore

$$\hat{\Gamma}_\varepsilon(v) \in S_\varepsilon(v).$$

The hull provides a geometric representation of the range and angular arrangement of the selected shell directions. Its shape may reveal directional concentration, separation, or uneven angular coverage. If $0 \in S_\varepsilon(v)$, then the selected shell directions admit at least one convex combination equal to zero.

The empirical shell-direction hull is related conceptually to generalized-gradient geometry, since both involve convex combinations of directional vectors. Nevertheless, it is not generally identical to the Clarke generalized gradient $\partial_C G_F(v)$. The latter characterizes limiting derivatives of the median-radius functional while $S_\varepsilon(v)$ is constructed directly from observations selected within a specified tolerance around the median-radius boundary.

### 4.2.2 Numerical illustration

A simulation study was conducted to investigate the geometric behaviour of the shell-direction hull under different distributional structures, including symmetric, skewed, bimodal, and contaminated distributions. For each setting, the active shell was identified using observations whose distances from the reference point were within approximately 10% of the empirical median radius. The corresponding directional vectors were then used to construct the convex hull.

Figure 5 presents the resulting shell-direction hulls for four simulated distributions.

For the bivariate Gaussian distribution (top-left panel), the shell directions are distributed around the reference point, producing a relatively balanced and approximately circular hull. This pattern is consistent with the symmetric structure of the underlying distribution.

For the skewed distribution (top-right panel), generated using an exponential component together with a Gaussian component, the hull exhibits an asymmetric configuration. This reflects differences in the angular arrangement of observations near the median-radius shell.

For the bimodal Gaussian mixture (bottom-left panel), the hull becomes elongated and approximately rectangular, reflecting the directional separation associated with the two concentration regions. This illustrates that the shell-direction hull can respond to multimodal structure even when the underlying distribution is not necessarily skewed.

For the contaminated Gaussian distribution (bottom-right panel), the hull remains approximately circular in the displayed example, with relatively limited deformation compared with the skewed and bimodal cases. This suggests that the selected median-radius shell may retain a relatively balanced directional configuration under the particular contamination setting considered.

Overall, Figure 5 illustrates how the shell-direction hull provides a geometric summary of local directional organization. Its shape complements the median-radius concentration landscape by describing angular features that are not represented by the scalar radius alone.

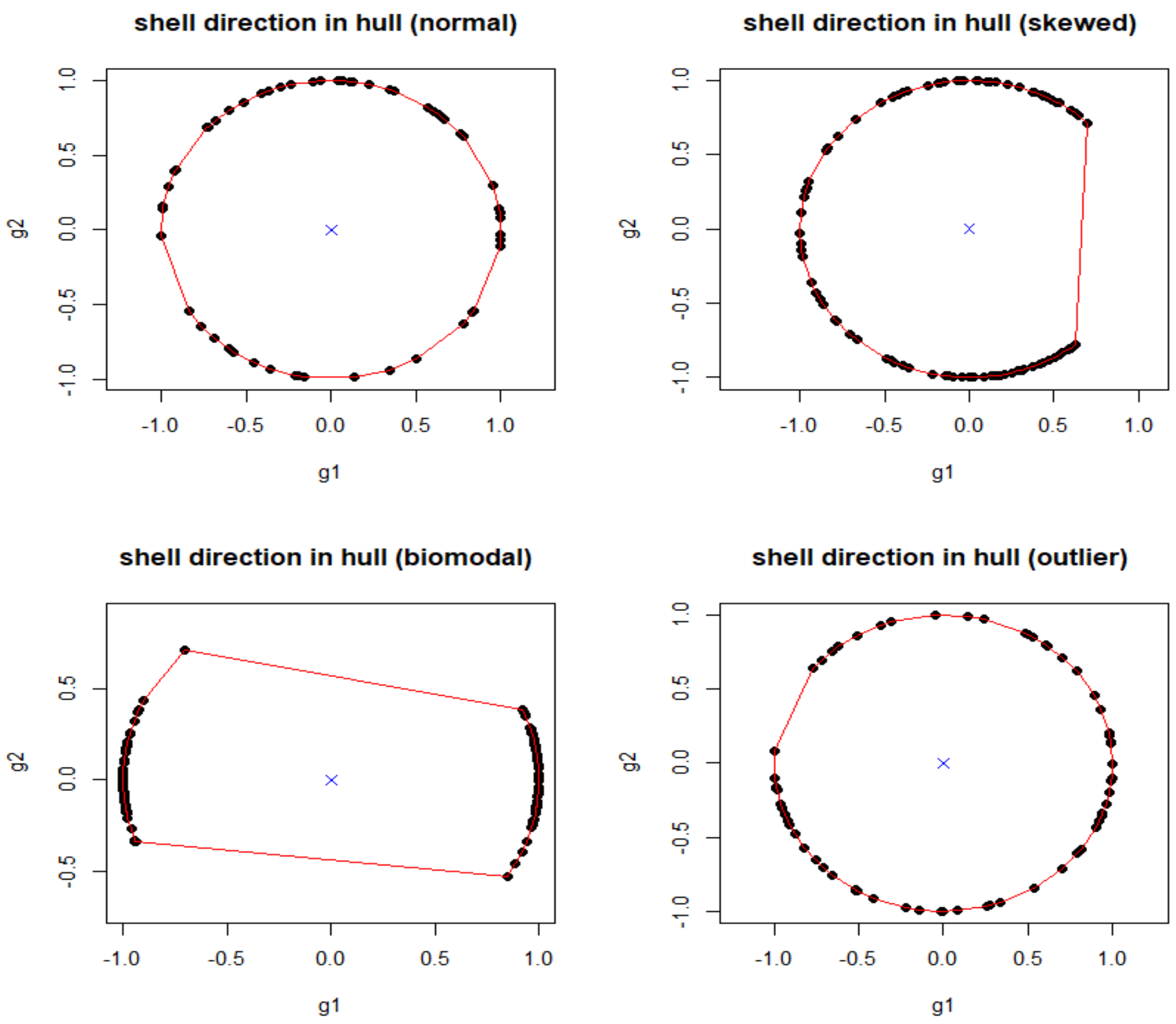

Figure 5. convexhull for simulated data from different distribution and $n = 1000$.

The proposed shell skewness measure and its associated convex hull therefore provide complementary descriptions of directional asymmetry

### 4.3 Comparison with classical directional skewness

Classical measures of multivariate skewness are mainly based on third-order moments and extend the idea of the univariate standardized third moment to higher dimensions. One of the most widely used measures is Mardia's multivariate skewness (Mardia et al., 1979), defined as

$$\beta_{1,d} = \mathbb{E}[\{(X-\mu)^{\top}\Sigma^{-1}(Y-\mu)\}^{3}],$$

where $X$ and $Y$ are independent copies of the random vector, $\mu$ is the mean, and $\Sigma$ is the covariance matrix. The sample version is

$$b_{1,d} = \frac{1}{n^2}\sum_{i=1}^{n}\sum_{j=1}^{n}\{(x_i-\bar{x})^{T}S^{-1}(x_j-\bar{x})\}^{3},$$

which aggregates third-order cross-moments across all directions and $\bar{x}$ and $S$ denote the sample mean and covariance matrix, respectively. This measure summarizes global multivariate asymmetry through third order standardized cross-moments.

Projection-based measures provide an alternative approach by examining skewness along selected directions. Following the directional framework discussed by (Ferreira & Steel, 2006), consider a unit vector $u \in \mathbb{R}^p$and define

$$\gamma(u) = \frac{\mathbb{E}[(u^{\mathrm{T}}X - u^{\mathrm{T}}\mu)^3]}{\{\mathbb{E}[(u^{\mathrm{T}}X - u^{\mathrm{T}}\mu)^2]\}^{3/2}},$$

An overall projection skewness index may be expressed as

$$\gamma_{\max} = \max_{\|u\|=1} \gamma(u).$$

This approach identifies directions exhibiting strong standardized third-moment asymmetry and therefore provides a more targeted, although still global, characterization of the distribution.

Despite their importance, moment-based and projection-based measures summarize asymmetry through distribution-wide quantities, which may obscure local geometric features. Recent approaches have consequently emphasized directional and spatial information. Projection pursuit methods (Zhang et al., 2024) investigate distributional structure through selected projections.

In contrast, the proposed shell skewness measure describes the directional organization of observations near the median-radius boundary. Let

$$\hat{\Gamma}_\varepsilon(v) = \frac{1}{|A_\varepsilon(v)|} \sum_{i \in A(v)} \frac{v - X_i}{\|v - X_i\|},$$

and its scalar magnitude is

$$S_\varepsilon(v) = \left\|\hat{\Gamma}_\varepsilon(v)\right\|.$$

Unlike classical skewness measures, the proposed measure does not require covariance estimation or third-order moments and is bounded between zero and one. Unlike Mardia's and projection-based skewness measures, it characterizes local directional imbalance around a specified reference point. Therefore, the three measures describe related but distinct aspects of multivariate asymmetry.

To investigate their sampling behaviour, a Monte Carlo simulation study is conducted under distributions with varying degrees of tail heaviness. Since the three statistics have different numerical scales and population interpretations, the comparison focuses on relative sampling variability rather than direct comparisons of their raw values, biases, or mean squared errors.

The simulation considers a bivariate normal distribution and multivariate Student $t$ distributions with degrees of freedom $\nu \in \{30,10,5,3,2\}$. For the normal reference, observations are generated from $N_2(0, I_2)$. For the student $t$ distributions, observations are generated according to

$$X_i = \frac{Z_i}{\sqrt{W_i/\nu}},$$

Where $Z_i \sim N_2(0, I_2), W_i \sim \chi^2_\nu$, and $Z_i$ and $W_i$ independent. This construction produces centrally symmetric distributions with increasingly heavy tails as $\nu$ decreases. Three sample sizes, $n \in \{50,100,200\}$ are considered with $B = 1000$ independent replications for each distribution and sample-size combination.

For each statistic $T_j$, $j \in \{M, P, S\}$ represents Mardia, projection and shell skewness, respectively, the Monte Carlo standard deviation is estimated across the $B$ replications. The primary comparison criterion is the relative sampling standard deviation,

$$RV_j(\nu, n) = \frac{SD_j(\nu, n)}{SD_j(N, n)},$$

Where $SD_j(N, n)$ is the corresponding standard deviation under the Gaussian reference distribution for the same sample size. Values close to one indicate variability comparable to the normal reference while larger values indicate increased sampling variability under heavy tails.
Table 1 reveals substantial differences in the sampling stability of the three skewness measures. As the degrees of freedom decrease, Mardia's skewness exhibits a pronounced increase in relative variability, particularly at larger sample sizes. Projection skewness also becomes increasingly variable, although the increase is less severe. In contrast, the proposed shell skewness measure maintains relative standard deviations close to the Gaussian reference across the considered distributions and sample sizes. For example, under the $t_2$ distribution with $n = 200$, the relative standard deviations are 639.113, 32.475, and 1.245 for Mardia, projection, and shell skewness, respectively.
The results demonstrate that the proposed shell skewness statistic exhibits comparatively stable sampling variability under symmetric heavy-tailed distributions. This behaviour is consistent with its bounded directional construction and its independence from covariance inversion and higher-order moments. Nevertheless, because all the simulated distributions are centrally symmetric, these findings concern stability rather than sensitivity to global asymmetry.

Table 1. relative sampling standard deviation ($RV$) of multivariate skewness measures

| $n$ | Distribution | Mardia | Projection | Shell |
|---|---|---|---|---|
| 50 | Normal | 1.000 | 1.000 | 1.000 |
| | $t_{30}$ | 1.608 | 1.313 | 0.968 |
| | $t_{10}$ | 2.863 | 1.888 | 0.993 |
| | $t_5$ | 10.504 | 3.759 | 1.036 |
| | $t_3$ | 22.344 | 5.716 | 1.012 |
| | $t_2$ | 38.714 | 7.668 | 1.055 |
| 100 | Normal | 1.000 | 1.000 | 1.000 |
| | $t_{30}$ | 1.298 | 1.173 | 1.025 |
| | $t_{10}$ | 4.262 | 2.106 | 1.034 |
| | $t_5$ | 20.360 | 4.926 | 1.019 |
| | $t_3$ | 71.692 | 10.272 | 1.112 |
| | $t_2$ | 130.924 | 14.771 | 1.186 |
| 200 | Normal | 1.000 | 1.000 | 1.000 |
| | $t_{30}$ | 1.573 | 1.313 | 1.088 |
| | $t_{10}$ | 6.673 | 2.671 | 1.066 |
| | $t_5$ | 60.657 | 8.070 | 1.093 |
| | $t_3$ | 277.346 | 20.640 | 1.212 |
| 200 | $t_2$ | 639.113 | 32.475 | 1.245 |

To complement the simulation, the geometric interpretation of shell skewness can be examined using a real bivariate dataset. The Old Faithful geyser data provide a suitable illustration because eruption duration and waiting time exhibit distinct concentration regions. Since the variables are measured on different scales, they are standardized before calculating Euclidean distances. Two complementary quantities summarize the directional geometry: the shell imbalance magnitude,

$$S_\varepsilon(c_F) = \left\| \frac{1}{A(c_F)} \sum_{i \in A} s_i \right\|,$$

and the distance from the origin to the directional convex hull,

$$D_S = \min_{z \in S} \|z\|.$$

That satisfies $0 \le D_S \le S \le 1$.

The shell imbalance magnitude measures the equal-weight directional imbalance while the convex-hull distance indicates whether the selected directions permit a balanced convex combination. In particular, $D_S = 0$ indicates that the origin belongs to the directional convex hull but does not necessarily imply that the average shell direction is zero.

Table 2. Median radius centre and shell direction geometry of the old faithful geyser data

| | | | |
|---|---|---|---|
| Number of observations | 272 | Active-shell percentage | 9.93% |
| Median-radius centre: standardized eruption | 0.714 | Shell imbalance vector: x | -0.049 |
| Median-radius centre: standardized waiting | 0.769 | Shell imbalance vector: y | 0.154 |
| Median radius | 0.648 | Shell imbalance magnitude | 0.161 |
| Relative shell tolerance | 0.10 | Convex-hull distance from origin | 0.000 |
| Absolute shell tolerance | 0.064 | Origin inside directional convex hull | Yes |
| Number of active-shell observations | 27 | Number of hull vertices | 24 |

Table 2 summarizes the median-radius and shell-direction characteristics of the standardized Old Faithful geyser data. The estimated median-radius centre is (0.7148, 0.7693), with a median radius of 0.6486. Using a relative shell tolerance of 10%, 27 observations (9.93%) are identified within the active shell. The shell imbalance magnitude is relatively small (0.1619). Furthermore, the convex-hull distance is zero  confirming the origin lies within the directional convex hull. Despite the presence of distinct concentration regions in the data, the results suggest that the active-shell directions exhibit a degree of directional balance. Thus, the shell imbalance and convex-hull measures provide complementary information about local directional organization rather than global distributional symmetry.

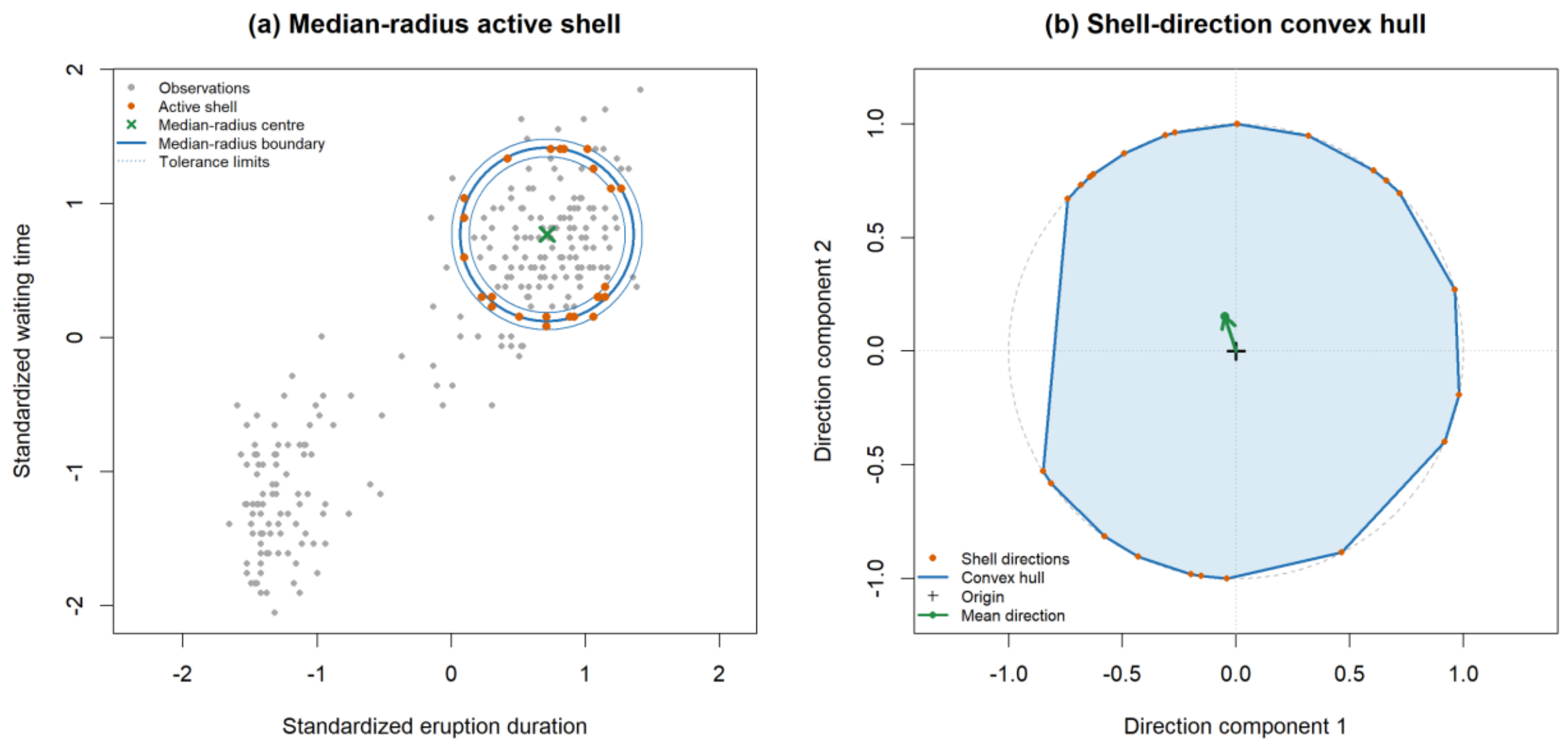

Fig 6. Median-radius shell geometry for the standardized Old Faithful geyser data.

Figure 6 demonstrates how the median-radius framework provides geometric information beyond a single global skewness coefficient. The active-shell directions describe the organization of observations around the selected reference centre while their convex hull and average direction summarize complementary aspects of local directional balance. These quantities should not be interpreted as direct tests of global symmetry or multimodality. Rather, they provide an additional geometric description that complements classical moment-based and projection-based skewness measures.

Therefore, the heavy-tailed simulation provides evidence of stable sampling variability while the convex-hull application demonstrates the interpretation of local shell-direction geometry. These findings support the use of shell skewness as a complementary, covariance-free diagnostic of multivariate distributional structure rather than a direct replacement for classical measures of global skewness.

# 5 Conclusion

This study has proposed and investigated the median-radius functional as a robust, covariance-free framework for describing multivariate centrality, radial concentration and directional structure. The functional is shown to be continuous, 1-Lipschitz and coercive. Unlike conventional convex distance functions, the median-radius functional need not be globally convex where this property arises naturally under multimodal distributions and allows the median-radius field to represent multiple concentration regions instead of imposing a single centre-outward structure. The standardized function $H(v)$ provides a relative measure of radial concentration. While Clarke generalized gradients provide a framework for analysing the functional at non-smooth points, an explicit derivative representation is obtained for smooth univariate distributions. In multivariate settings, the median-radius construction is equivariant under translations, orthogonal transformations and common scaling without requiring covariance estimation or matrix inversion. The associated shell-direction convex hull and directional imbalance measure provide additional information about the local organization of observations around the median-radius boundary. These geometric quantities complement classical multivariate skewness measures by describing local directional structure rather than relying on global third-order moments.

The numerical applications illustrate the potential advantages of the proposed framework under non-elliptical, multimodal, and high-dimensional data configurations. Comparisons with classical and robust Mahalanobis distances demonstrate how median-radius contours can represent concentration patterns that are not necessarily captured by elliptical distance contours. Furthermore, the Monte Carlo comparison under symmetric heavy-tailed distributions indicates that shell skewness maintains relatively stable sampling variability compared with Mardia's and projection-based skewness measures. The real-data shell analysis further illustrates how the directional convex hull and shell imbalance magnitude can provide complementary descriptions of local directional balance.

Despite its geometric flexibility and covariance-free construction, the proposed approach has some limitations. The median-radius functional may overlook small-scale distributional features because it focuses on the radius enclosing half of the probability mass. Its nonconvexity can also complicate numerical optimization especially when multiple local minima are present.

Future research may investigate the asymptotic properties and influence functions of median-radius centres and shell imbalance measures. Further work could also examine formal inference procedures for detecting multimodality and directional asymmetry.

## Appendix

**Proof of theorem 1:**
From the triangle inequality

$$\|X - v\| \leq \|X - w\| + \|v - w\|.$$

Quantile monotonicity gives

$$G_F(v) \leq G_F(w) + \|v - w\|.$$

Interchanging $v$ and $w$,

$$G_F(w) \leq G_F(v) + \|v - w\|.$$

Hence,

$$|G_F(v) - G_F(w)| \leq \|v - w\|$$

**Proof of theorem 2:**
Choose $M > 0$ such that

$$P(\|X\| \leq M) > 0.5$$

For observations satisfying $\|X\| \leq M$,

$$\|X - v\| \geq \|v\| - \|X\| \geq \|v\| - M.$$

Since this event contains more than half the probability mass,

$$G_F(v) \geq \|v\| - M.$$

Therefore,

$$G_F(v) \to \infty.$$

Hence, continuity plus coercivity guarantees that the minimum is attained.

**Proof of proposition:**
Consider the continuous mixture distribution

$$X \sim 0.5U(-2,-1) + 05U(1,2).$$

Its density is

$$f(x) = \begin{cases} 0.5, & -2 \le x \le -1, \\ 0.5, & 1 \le x \le 2, \\ 0, & \text{otherwise.} \end{cases}$$

Using median radius

$$G_F(v) = \inf\{r \ge 0: P(|X - v| \le r) \ge 0.5\}.$$

At $v = -1.5$, the interval $[-2,-1]$ contains exactly half of the probability mass. Hence, $G_F(-1.5) = 0.5$. By symmetry $G_F(1.5) = 0.5$.
At $v = 0$, the interval $[-r, r]$ contains no probability mass for $r < 1$ and $1 \le r \le 2$, $P(|X| \le r) = r - 1$. Therefore, $G_F(0) = 1.5 > 0.5\big[G_F\big(-1.5 + G_F(1.5)\big)\big] = 0.5$.
This violates the midpoint convexity inequality. Therefore, $G_F(v)$ is not convex in general.

**Proof of theorem 3:**
Since

$$F\big(v + G(v)\big) - F\big(v - G(v)\big) = 0.5.$$

Differentiate

$$f\big(v + G(v)\big)\big(1 + G'(v)\big) - f\big(v - G(v)\big)\big(1 - G'(v)\big) = 0,$$

Which gives

$$G'(v) = \frac{f\big(v - G(v)\big) - f\big(v + G(v)\big)}{f\big(v - G(v)\big) + f\big(v + G(v)\big)}.$$

**Proof corollary:**
This follows using

$$|a - b| \le a + b,$$

For nonnegative $a, b$.

**Proof Theorem 4:**
Since $Y = a + bQX$, we have

$$\|Y - (a + bQv)\| = b\|Q(X - v)\| = b\|X - v\|,$$

Where the last equality follows from orthogonality of $Q$.
Taking medians and using the positive scale equivariance of quantile gives

$$G_Y(a + bQv) = Med(b\|X - v\|) = bG_X(v).$$

Because the transformation $v \mapsto a + bQv$ is bijective, minimizing both sides yields $C_Y = a + bQC_X$ and $g_Y = bg_X$. Using these results

$$H_Y(a + bQv) = \frac{bG_X(v)}{bg_X} = H_X(v).$$